\documentstyle[12pt,aaspp]{article}
\def\ltsima{$\; \buildrel < \over \sim \;$}
\def\lsim{\lower.5ex\hbox{\ltsima}}
\def\gtsima{$\; \buildrel > \over \sim \;$}
\def\gsim{\lower.5ex\hbox{\gtsima}}
\begin{document}
\title{Self-Similar Evolution of Gravitational Clustering II: 
N-Body Simulations of the $n=-2$ Spectrum}

\author{Bhuvnesh Jain\altaffilmark{1}and Edmund Bertschinger\altaffilmark{2}}
\affil{\altaffilmark{1}Department of Physics \& Astronomy, Johns Hopkins
University, Baltimore, MD 21218. Email: bjain@pha.jhu.edu}
\affil{\altaffilmark{2}Department of Physics, MIT, Cambridge, MA 02139. 
Email: edbert@arcturus.mit.edu}

\begin{abstract}
The power spectrum $P(k)\propto k^n$ with $n=-2$ is close to
the shape of the measured galaxy spectrum on small scales. 
Unfortunately this spectrum has proven rather difficult
to simulate. Further, 2-dimensional simulations have suggested
a breakdown of self-similar scaling for spectra with $n<-1$ due 
to divergent contributions from the coupling of long wave modes.  
This paper is the second (numerical) part of our investigation 
into nonlinear gravitational clustering of scale-free spectra,
in particular to test the scaling of the $n=-2$ spectrum.

Using high-resolution N-body simulations we find that the $n=-2$ 
power spectrum displays self-similar scaling. The phase shift of 
Fourier modes of the density show a dual scaling, 
self-similar scaling at early times and a scaling driven by 
the bulk velocity at late times. The second scaling was shown
analytically to be a kinematical effect which does not affect the 
growth of clustering. Thus our analytical and N-body results verify that 
self-similarity in gravitational clustering holds for $-3<n<1$. 
The N-body spectrum is also compared with analytic fitting formulae, which
are found to slightly underestimate the power in the nonlinear regime.
The asymptotic shape of the spectrum at high-$k$ is a power law  
with the same slope as predicted by the stable clustering hypothesis. 
\end{abstract}


\section{Introduction}

The self-similar scaling of density perturbations with
scale free initial conditions in a spatially flat universe is a
useful theoretical tool in studying structure formation. It has been
widely used to study gravitational clustering in cosmology and has been
tested by several studies using N-body simulations. However 
self-similar scaling for scale free initial 
spectra $P_{\rm lin}(k)\propto k^n$ has not been adequately demonstrated for 
$n<-1$ because the requirements of dynamic range get increasingly difficult
to meet as $n$ gets smaller. Indeed 
results of some two dimensional studies suggest a breaking of self-similar
scaling for $n=-2$ in three dimensions. 
Analytical analyses have been limited to the observation
that the linear peculiar velocity field diverges for $n<-1$, 
but the linear density contrast does not diverge provided $n>-3$.  This 
would suggest that while
there may be formal problems with establishing self-similarity for $n<-1$, 
in practice it should hold as long as $n>-3$. 

In an earlier paper (Jain \& Bertschinger 1996 -- Paper I), we have
analyzed the dynamics of the coupling of long wave modes by 
analytical techniques to address the question of whether self-similar 
scaling is broken for $n<-1$. On the basis of the nonlinear
growth of density perturbations we  concluded that 
self-similar scaling does not break down provided $n>-3$, consistent
with the linear theory expectation. We also identified statistical
measures which would scale differently from self-similar scaling
owing to the kinematic effect of large-scale bulk flows. These
measures are related to the phase of the Fourier space density field, 
but do not affect the growth of its amplitude. 

In this paper we
examine the scaling behavior of the $n=-2$ spectrum using high 
resolution three dimensional N-body simulations. 
Testing self-similarity using simulations
is crucial because the analytical demonstration of a scale-free
dynamics provides no guarantee that the similarity solution will 
actually hold, starting from some general initial conditions. 
Moreover, the analytical
results of Paper I relied on an assumption about the initial fields,
as well as on using the fluid limit approximation. The N-body 
simulations do not rely on these approximations, though as we 
shall see their finite resolution introduces other departures 
from exact self-similarity. Hence by combining
our analytical and numerical results, we attempt to provide a complete
analysis of the scaling behavior of scale free spectra for $-3<n<1$. 

We shall consider the similarity properties of gravitational dynamics in a 
zero-pressure Einstein-de Sitter cosmology. An Einstein-de Sitter universe 
refers to the model with the cosmological density parameter 
$\Omega=\Omega_{matter}=1$ and zero cosmological constant, 
so that the universe is spatially flat.
The gravitational interaction also does not pick a special length scale.
Further let the initial power spectrum be a power law, $P_{\rm lin}(k)\propto k^n$, over
length scales of interest. In such a case there is no preferred length
scale in the universe, and the evolution of structure is expected to be
self-similar in time. 

The resulting self-similar scaling of characteristic length scales $x$, and 
wavenumber scales $k$ is (Peebles 1980, Section 73):
\begin{equation}
x_{ss}(t)\propto a(t)^{2/(3+n)}\,\,\,\, ;\,\,\,\, k_{ss}(t)\sim 
x_{ss}(t)^{-1} \propto a(t)^{-2/(3+n)}, 
\label{ss}\end{equation}%(4.2)
where $a(t)$ is the expansion scale factor. The above scaling is
most simply derived by requiring consistency with the growth of
the rms linear density contrast, smoothed on a scale
$x_{ss}=k_{ss}^{-1}$, and related to the power spectrum as:
\begin{equation}
\left({\delta\rho\over\rho}\right)^2(x_{ss},t)\, =\, \int d^3k \,a^2\,P_{\rm lin}(k)
\, W^2(kx_{ss}) \,\simeq\, 4\pi \,a^2\, k_{ss}^{3+n}\, ,
\label{drho}\end{equation}
where $W(kx)$ is the smoothing window function. 
Requiring that the scale at which $({\delta\rho/\rho})=1$ scales
self-similarly gives the result in equation (\ref{ss}).

Early studies of self-similar evolution in cosmology include those
of Peebles (1974); Press \& Schechter (1974); Davis \& Peebles (1977);
and Efstathiou \& Eastwood (1981). 
Efstathiou et al. (1988) tested self-similar
scaling in N-body simulations of scale free spectra with $n=-2,-1,0,1$. 
They examined the scaling of the correlation function $\xi(x,t)$,
and of the multiplicity function describing the distribution of bound 
objects. They verified the predicted scaling for both statistics, and
found consistency with the picture of hierarchical formation of nonlinear
structure on increasingly large length scales. Their results for 
$n=-2$ did not match with the self-similar scaling as well as 
the other cases. Bertschinger \& Gelb (1991) used better resolution 
simulations to address these questions and also found similar results. 
These authors concluded that the reason for the weakness of the $n=-2$ 
results was the finite size of their simulation box, as the $n=-2$ case
has more power on large scales and therefore requires
a larger box-size to approximate the infinite volume limit with the 
same accuracy as larger values of $n$. 

More recently Lacey \& Cole (1994) have examined the self-similar scaling of
the number density of nonlinear clumps for scale-free spectra. Their results
indicate that self-similar scaling works reasonably well for the
statistics they measure, even for $n=-2$. Jain, Mo \& White (1995)
and Padmanabhan et al. (1995) reach the
same conclusion for the correlation function and power spectrum. 
Colombi, Bouchet \& Hernquist (1995) have verified self-similarity 
for higher moments of the density. However, in all the above studies
the self-similar scaling of a particular statistic is tested without making
a comparison with alternate scalings. As demonstrated in Section 4,
this makes it difficult to distinguish the effect of limited numerical
resolution from a real breakdown of self-similar scaling. 
%We will demonstrate that even with the current 
%state-of-the-art  simulations it
%is extremely difficult to use an averaged quantity like the power 
%spectrum to critically test self-similar scaling for $n=-2$. 

The N-body results of Ryden \& Gramann (1991), and Gramann (1992)
suggest that the $n=-2$ case is different from $n\ge -1$ for a fundamental 
reason. They studied $n=-1$ simulations in two dimensions, which are the 
analog of $n=-2$ in three dimensions,
and examined the scaling of the phase (Ryden \&
Gramann 1991), and then both phase and amplitude (Gramann 1992)
of the Fourier transform of the density field. The scaling was found to be
different from the standard self-similar scaling. 
Characteristic wavenumber scales, instead of following the self-similar 
scaling, given in two dimensions by 
$\, k_{ss}(t) \propto a(t)^{-2/(2+n)}\propto a(t)^{-2}$, showed the 
scaling $k \propto a(t)^{-1}$. 
%They pointed out that this scaling would be 
%obtained if the linear bulk velocity field were used to define characteristic
%scales, as opposed to the conventional choice of using the rms 
%$\delta\rho/\rho$. 
Other studies in two dimensions also suggest that 
a transition in nonlinear evolution occurs at $n=-1$ (Klypin \& Melott 1992, 
and references therein). 
Motivated by Gramann's results we had re-examined the 
$n=-2$ simulation presented in Bertschinger \& Gelb (1991) and found
that the results were ambiguous, and that a bigger simulation would be
needed to provide a definitive answer. 

In this paper we use the $n=-2$ case as representative of the 
range $-3<n<-1$ and  use high resolution N-body simulations to test
for self-similar scaling. Section 2 provides the analytical background
and motivates the need to test the scaling of spectra with $n<-1$. 
The N-body simulations used to test self-similar scaling are 
described in Section 3, and effects of finite numerical resolution
are discussed. 
In Section 4 results for the scaling of the power spectrum are presented
and self-similar scaling is compared with an alternate scaling.
In Section 5 a different statistic is used which
measures the scaling of the amplitude as well as the phase of 
the Fourier space density. We discuss subtleties in measuring
the evolution of the phase, and provide a kinematical interpretation
of the results for the phase shift. We conclude in Section 6
by discussing the results of this paper and Paper I. 

\section{Analytical Background}

This section summarizes the formalism and motivation presented in
Section 2 of Paper I. We use comoving 
coordinates $\vec x$ and conformal time $d\tau=dt/a(t)$
to write the nonrelativistic cosmological fluid equations as
\begin{mathletters}
\begin{equation}
{\partial\delta\over\partial\tau}+\vec\nabla\cdot[(1+\delta)\vec v\,]=0 \ ,
%\eqno(1a)
\label{continuity}
\end{equation}
\begin{equation}
{\partial\vec v\over\partial\tau}+\left(\vec v\cdot
\vec\nabla\right)\vec v=-{\dot a\over a}\,\vec v-\vec\nabla\phi\ ,
%\eqno(1b)
\label{euler}%\eqnum{1b}
\end{equation}
\begin{equation}
\nabla^2\phi=4\pi Ga^2\bar\rho\delta\ ,
%\eqno(1c)
\label{poisson}%\eqnum{1c}
\end{equation}
\label{eqs1}\end{mathletters}
where $\dot a\equiv da/d\tau$, $\delta(\vec x,\tau)=(\rho(\vec x,\tau)-
\bar\rho(\tau))/\bar\rho(\tau)$, $\phi(\vec x,\tau)$ is the 
perturbed gravitational potential, and
$\vec v(\vec x,\tau)\equiv d\vec x/d\tau$ is the proper 
peculiar velocity. 
We assume an Einstein-de Sitter ($\Omega=1$) universe, with $a\propto
t^{2/3}\propto\tau^2$. We will also assume that the initial
(linear) density fluctuation field is a Gaussian random field.

The fluid equations (\ref{eqs1}) are strictly valid only on scales
large compared to the nonlinear clustering scale. On scales much
smaller than the nonlinear scale, the intersection of particle 
trajectories leads to a  complicated, anisotropic stress tensor
and thus invalidates a fluid description which requires the 
density and velocity to be smooth, single-valued functions of position. 
Whereas the analytical analysis of Paper I used the zero-pressure fluid
approximation and the continuum limit, the N-body simulations do 
not as they integrate the equations of motion for the particle
trajectories. 

To quantify the amplitude of fluctuations on various scales it is preferable
to work with the Fourier transform of the density fluctuation field,
which we define as
\begin{equation}\hat\delta(\vec k,\tau)=\int{d^3x\over(2\pi)^3}\,
e^{-i\vec k\cdot\vec x}\,\delta(\vec x,\tau)\ .
%\eqno(2)
\label{deltak}\end{equation}
The power spectrum
(power spectral density) of $\delta(\vec x,\tau)$ is defined by
the ensemble average two-point function,
\begin{equation}\langle\hat\delta(\vec k_1,\tau)\,
\hat\delta(\vec k_2,\tau)\rangle=
P(k_1,\tau)\,\delta_{\rm D}(\vec k_1+\vec k_2)\ ,
%\eqno(3)
\label{pk}\end{equation}
where $\delta_{\rm D}$ is the Dirac delta function, required for a spatially
homogeneous random density field.  For a homogeneous and isotropic random
field the power spectrum depends only on the magnitude of the wavevector.
The contribution to the variance of $\delta(\vec x,\tau)$ from waves in the
wavevector volume element $d^3k$ is $P(k,\tau)d^3k$.
The autocorrelation function is defined as
$\xi(x,\tau)=\langle \delta(\vec
x_1,\tau)\delta(\vec x_2,\tau)\rangle$, where $|\vec x_1-\vec x_2|=x$. It can
be easily verified that $\xi(x,\tau)$ is the Fourier transform of $P(k,\tau)$. 

The analytical approach developed in Paper I can now be summarized
as follows. For a scale free
initial spectrum $P_{\rm lin}(k)\propto k^n$, the rms density contrast 
in linear theory is given by equation (\ref{drho}). As $k\to 0$ the
window function generically (e.g. the top-hat or Gaussian window functions) 
has the limiting form $W(kx)\to 1$. Hence for $n\leq-3$ 
the linear rms density contrast on any scale receives divergent contributions
from the $k\to0$ part of the spectrum. Analogous to the rms density 
contrast one can define the bulk flow velocity, which is the
average, smoothed peculiar velocity on scale $x$. It is obtained
by using the peculiar velocity power spectrum $P_{{\rm lin}\, v}$ in equation
(\ref{drho}) instead of $P_{\rm lin}$. 
From equation (\ref{continuity}) it follows that
$\dot\delta=-\vec \nabla\cdot \vec v$ in linear theory, or $P_{{\rm lin} v}(k)=
\dot a^2/a^2\, P_{\rm lin}(k)/k^2$ . Thus the bulk velocity is given by
\begin{equation}
v_b^2(x,\tau)=
%\int d^3 k P_{{\rm lin}\, v}(k,\tau) W^2(kx)=
\dot a^2 \, 4 \pi \int d k P_{\rm lin}(k) W^2(kx). 
\label{vbulk}\end{equation} %4.8
For $n<-1$, $v_b(x,\tau)$ diverges due to the $k\to 0$
contribution. Since the nonlinear fluid equations couple
the density and velocity, this divergence manifests itself in the
nonlinear terms in the fluid equations. 

If this divergence were to have a dynamical
influence in the growth of the density contrast it would lead to a
breakdown of self-similar scaling, because it introduces a new scale
in the system. Setting $v_b(x)={\rm constant}\le c$, with a finite 
lower limit in 
the integral in equation (\ref{vbulk}) gives the scaling $x\propto a$
for all $n<-1$. The standard self-similar scaling of equation
(\ref{ss}) then holds only if $n>-1$. 
This, coupled with the numerical results of
Gramann (1992) described in Section 1, provided the motivation for
our analytical examination of spectra with $n<-1$ in Paper I. 
The results of Paper I are summarized in Section 5.
In the following sections we will analyze the
scaling of the $n=-2$ spectrum in N-body simulations to test 
whether it follows the velocity scaling $x\propto a$ or the 
self-similar scaling of equation (\ref{ss}): $x\propto a^2$.

%Fourier transforming equations (\ref{eqs1}) gives:
%\begin{mathletters}\begin{equation}
%{\partial\delta\over\partial\tau}+ \theta=-
%\int\!d^3k_1\,{\vec k\cdot\vec k_1\over k_1^2}\,
% \theta(\vec k_1,\tau)\,\delta(\vec k - \vec k_1,\tau)\ ,
%\label{ft1}\end{equation}%(4.3a)
%\begin{equation}
%{\partial \theta\over\partial\tau}+{\dot a\over a}\, \theta+
%{6\over\tau^2}\,\delta=- \int\!d^3k_1\,
%k^2 {\vec k_1\cdot(\vec k-\vec k_1)\over2k_1^2 \vert \vec k-\vec k_1\vert^2}
%\, \theta(\vec k_1,\tau)\, \theta(\vec k-\vec k_1,\tau)\ .
%\label{ft2}\end{equation}%(4.3b)
%\label{fts}\end{mathletters}
%The fields on the left-hand side are all functions of $\vec k$ and $\tau$.

\section{The N-body Simulations}

N-body simulations provide a powerful means for testing the 
self-similar scaling of scale free spectra. The deeply nonlinear regime is 
accessible in these
simulations, thus offering the possibility of measuring the complete
similarity solution. N-body simulations have limited dynamic range, but they
do not rely on any approximations of the kind made in Paper I. Since the
equations of motion for individual particles are integrated, the fluid
limit approximation is not required either. 
Therefore they provide a complementary technique to the analytic
approaches of Paper I. 

The simulations used for the self-similar scaling analysis
are three dimensional particle-particle/particle-mesh (P$^3$M) simulations. 
The larger of our $n=-2$ simulations had $256^3$ particles and 
a Plummer force softening parameter $\epsilon= 1/5120\, L$, where 
$L$ is the box-length. It was performed on the CM-5 supercomputer
at the National Center for Supercomputing Applications using the parallel code
of Ferrell \& Bertschinger (1994, 1995). 
This simulation, labeled A in Table 1, is used
for the main results of this paper. To test the effects of numerical 
resolution at high-$k$ we have compared the results with a smaller P$^3$M
simulation (labeled B), performed on a workstation,
which used half as many particles and twice the force softening
length. This simulation used a variable PM mesh as shown in table 1. 
We have also used a set of three PM simulations (labeled C),
performed on the CM-5, to
estimate the effects of statistical fluctuations at low-$k$. 
Table 1 gives the relevant parameters of all the simulations used. 

\begin{table}
\caption{Parameters of the N-body simulations}
\medskip
\begin{tabular}{|l|c|c|c|c|c|}			\hline

Simulation & {\bf $k_{nl}/(2\pi/L)$}  &{\bf $N_{part}$ } & 
Softening $\epsilon/L$ & PM Mesh & Timesteps \\   \hline \hline

A: $n=-2$, P$^3$M & $5250-7.25$ & $256^3$ & $1/5120$ & $512^3$ & $1331$\\ 
\hline

B: $n=-2$, P$^3$M & $2060-4.02$ & $128^3$ & $1/2560$ & $256^3-432^3$ &
$1007$\\ 
\hline

C: $n=-2$, PM   & $5250-7.25$ & $256^3$ & ---       & $512^3$  &$692$\\ 
\hline
\end{tabular}
\label{Table}
\end{table}

%The $n=0$ simulation was performed by Simon White. It is also a 
%particle-particle/particle-mesh simulation with $100^3$ particles
%and $\epsilon=L/2500$. 
%Data were available for $9$ time outputs, with the scale factor increasing
%by a factor of $\simeq 1.6$ between successive outputs (except for the 
%first two). 
%The $n=0$ simulation is also a particle-mesh simulation with $256^3$
%particles.
%For the $n=0$ simulation $11$ time outputs were analyzed with $a(\tau)$
%increasing by a factor of $2$ between successive outputs. 
%This simulation is used as a control case to test the accuracy of the 
%results for the $n=-2$ simulation. Since $n=0$ does not suffer from 
%potential long wave divergences, the standard 
%self-similar scaling is expected to hold, and has in fact 
%been convincingly demonstrated in earlier work. 
%Hence the results 
%of this simulation provide a good indication of the degree of accuracy
%with which the N-body simulations show the similarity solution.

The initial positions and velocities for the particles were generated
using the Zel'dovich approximation, and assumed the linear power 
spectrum and Gaussian statistics. Outputs of the particle
positions were stored at time intervals corresponding to 
an increase of $a$ by a factor of $2^{1/4}$. Energy conservation
was better than $10^{-4}$  of the total gravitational energy 
at all times for simulation A and was similar for the other simulations. 

At each of these output times the particle positions were interpolated onto 
a $1024^3$ grid using the triangular-shaped-cloud (TSC) scheme (see
Gelb \& Bertschinger 1994 for details) to
get the real space density. The density field $\delta(\vec x,a)$ 
was then fast Fourier transformed to get 
$\hat \delta(\vec k,a)$. $\hat \delta(\vec k,a)$
is a complex number at each $\vec k$, and is therefore represented by a 
real and imaginary part. The values of $\vec k$ are represented by a 
three dimensional vector $(k_x,k_y,k_z)$, with each of $k_x,k_y$ and $k_z$ 
being integers ranging from $-511$ to $512$. In the remainder of
this section the magnitude of $\vec k$ will be given in units of $2\pi/L$, 
so that the modes with wavelength equal to $L$ have $k=1$.
%Since $\hat \delta(\vec k)=\hat 
%\delta^*(- \vec k)$, only one half 
%of the values of $\hat \delta(\vec k)$ are independent; thus, for example, 
%the half of Fourier space with $k_z < 0$ does not contain any independent
%values of $\hat \delta(\vec k)$ if all the values for $k_z > 0$ are known. 
The power spectrum was computed as the mean of $|\hat\delta(\vec k,a)|^2$
in bins of $k$ with width $\delta k=1$. The power spectrum was
corrected for the effects of shot noise and of convolution with the 
TSC interpolating window used to compute the real space density. 
Figure 1 shows the power spectrum for simulation A at five different
values of $a$, corresponding to values of the nonlinear wavenumber
(defined in equation (\ref{knl}) below) $k_{nl}(a)=58, 29, 14.5,
7.25$. Also shown as the dotted line is the linear spectrum at the same times. 

In an N-body simulation there are many departures
from the idealization of an infinite, continuous fluid. 
The limitations common to any
N-body simulation are the finite size of the box and the discrete nature
of particles. Additional scales are introduced 
due to the PM mesh, force softening and the density
interpolation grid. As discussed in Section 1 of Paper I, the presence
of these scales is not a fundamental
drawback; it only means that one must be careful to ensure that the range of
scales used to study the scaling in time is sufficient to allow for 
intermediate asymptotic self-similarity to set in. In the following
discussion we shall focus on the largest of our simulations, labeled
A in Table 1. 

An important scale that we shall refer to is the nonlinear scale, 
at which the fluctuation amplitude is $\sim 1$. We define a nonlinear
wavenumber as:
\begin{equation}
\int_0^{k_{nl}} d^3k \, P_{\rm lin}(a,k)\, =\, 1\, .
\label{knl}\end{equation}
For $P_{\rm lin}(a,k)=A a^2 k^{-2}$, equation (\ref{knl}) gives $k_{nl}(a)=
(4\pi A a^2)^{-1}$. Thus $k_{nl}(a)$ becomes smaller at late times
since the nonlinear length scale increases with time. For scale
free spectra $k_{nl}(a)$ provides a useful time variable, for example,
in comparing simulations with different starting amplitudes. 
We now turn to the main numerical limitations in our simulations. 

(i) {\it Initial amplitude.} The initial amplitude used in our $n=-2$ 
simulations A and C is such that at $a=1$ the dimensionless power on
the Nyquist frequency of the particle grid, $k=128$ is $4 \pi k^3
P(k)=0.024$. This corresponds to having $k_{nl}=5250$ at the
initial time. The $n=-2$ 
spectrum needs to be started with very small initial amplitude in
order for clustering on small scales to develop self-similarly. 
As pointed out by Lacey \& Cole (1994) the standard prescription of
Efstathiou et al. (1985) of having power on the Nyquist frequency of
the particle grid equal to the white noise level is too large and 
produces significant departures from self-similar evolution. 

(ii) {\it Force softening.} Force softening is implemented in
computing short range forces in the particle-particle part of 
the P$^3$M simulations by using the Plummer force law: 
$F(r)=Gm^2/(r^2+\epsilon^2)$. The parameter $\epsilon$ is given
in Table 1 for the various P$^3$M simulations. The distance at which
$r^2F/(Gm^2)=1/2$ is $\simeq 1.3 \epsilon$. This distance is 
$\simeq L/4000$ for simulation A: it is smaller by almost a
factor of $4$ than the main limiting factor on small scales, the spacing 
of the $1024^3$ grid used for interpolating the density
(as shown below in point (iii), the grid 
effects are in fact significant up to twice the grid spacing). This is 
important, since the force softening affects the dynamical evolution 
and can therefore artificially suppress the clustering on scales
several times larger than $\epsilon$. 
%Note that for the P$^3$M simulations the PM mesh 
%is used for computing the long-range forces, but does not affect
%the accuracy of forces on small scales. 
In addition, the finite number of particles and the finite
size of the timesteps used limit the numerical resolution on small
scales. For the parameters in our simulations these affect
the spectrum on about the same limiting scale as the force softening. 
Hence their effect is also far less important than that of the FFT grid, 
aside from the PM simulations whose force softening was relatively
large. 

(iii) {\it FFT grid.} The density field was fast Fourier transformed 
to get $\hat \delta(\vec k)$. This involves interpolation
on to a grid to get the real space density field, and thus imposes
a minimum length scale below which power cannot be measured. Figure
2 shows a comparison of the power spectrum computed for simulation B
using two different grid sizes: $512^3$ and $1024^3$. Both spectra
have been corrected for the effects of convolution with the TSC window.
It shows that
the power for $k$ up to $1/2$ the Nyquist frequency of the FFT grid
is accurate to better than $20\%$. 
However this is only a consistency
check,
% --- there is in fact a suppression of the power at $k$ slightly
%lower than $1/2$ the Nyquist frequency.
as a conservative estimate, we therefore use a maximum $k=256$
for the $1024^3$ interpolation and FFT grid in our self-similarity
analysis. 

(iv) {\it Statistical fluctuations at low-$k$.} At low wavenumbers
the finite number of modes present in a given bin in $k$ causes 
statistical fluctuations in the power spectrum. This is evident in 
Figure 1 which shows that at $k=1$ and $2$ 
there are significant differences between $P(k)$ and $P_{\rm lin}(k)$
even at early times. By checking the power spectra of the three
different PM runs of simulation C we found that as the simulation
evolved the power in realizations with lower initial power at 
$k=1$ and $2$ was suppressed out to $k\simeq 5$ due to nonlinear 
mode coupling. For the latest time output used in our analysis,
at $k\geq 5$ these fluctuations are smaller than $20\%$. For the
results shown in Section 4 we plot the power spectra from $k=3-256$. 

(v) {\it Range of $k_{nl}(a)$.} At late times the nonlinear
wavenumber $k_{nl}(a)$ approaches the size of the box. This
sets a maximum value of $a$ beyond which
the absence of modes with wavelength larger than the box-size inhibits
the  growth of the spectrum at late times, because the
coupling of these modes would have enhanced the power at high-$k$
(Jain \& Bertschinger 1994). 
To test this effect, simulation B was 
evolved  to a maximum $a$ corresponding to $k_{nl}=2$. 
Using this simulation we find that for 
the $n=-2$ spectrum the power at $k\lsim 10$ is indeed underestimated for 
$k_{nl}(a)\lsim 4$. At later times, for example with $k_{nl}(a)=2.8$,
the power spectrum at $k$ extending up to $k\simeq 100$ is significantly 
suppressed relative to the amplitude required to satisfy
self-similarity. This problem is particularly severe for the $n=-2$ 
spectrum because of the large amount of power on low-$k$
relative to spectra with larger $n$, which in turns leads to a
stronger nonlinear coupling with high-$k$ modes. 
It is the principal reason for the requirement of a large dynamic
range to test the scaling properties of this spectrum. 

Our choice of $k_{nl}(a_{max})=7.25$ for 
simulation A is adequate to get dynamically accurate power spectra
at low-$k$, and also provides a sufficient range in $a$ to test
self-similar evolution. 
At small $a$ we are limited by the fact that the shot noise amplitude
which we subtract from the measured power spectrum 
exceeds the measured spectrum at high-$k$ for 
$a<8$ ($k_{nl}=82$). This determines the earliest output used in our 
analysis -- for simulation A this had $k_{nl}(a_{min})=58$.

(vi) {\it Accuracy of PM simulations.}
A final point, regarding the adequacy of PM simulations to follow
the scaling of the $n=-2$ spectrum is made in Figure 3, where the
power spectrum of the P$^3$M simulation A is compared with the 
PM simulation C. These simulations have identical 
number of particles, the same initial Fourier modes, and are started 
with the same $k_{nl}(a)$. Both spectra are plotted upto 1/2 the
Nyquist frequency of the FFT grid, and can therefore be legitimately
compared over the full range of $k$ for which they overlap. 
The figure shows that at late times the power 
in the PM simulation is suppressed at wavenumbers significantly
smaller than the Nyquist frequency of the PM grid, $k=512$. At the
last output time shown, corresponding to $k_{nl}(a)=7.25$, the 
power at $k>50$ is suppressed by $>20\%$.
The suppression increases to $>50\%$ at $k>80$, i.e. at $k >1/6$th 
the Nyquist frequency of the PM grid. Thus the commonly cited 
limit of the spatial resolution of PM simulations as being
less than twice the grid spacing considerably overestimates the spatial 
resolution. Our results are at odds with previous tests of PM 
simulations in the literature (e.g. Splinter, Melott \& Shandarin 1997).
The difference in these tests is the statistical measures
used as well as the degree of nonlinearity, since the discrepancy 
appears to occur only in the regime of very
high over-densities (the last output time shown).

\section{Scaling of the Power Spectrum}

The self-similar solution for the power spectrum and
autocorrelation function is (Peebles 1980, Section 73), 
\begin{equation}
P(k,\tau)=a^{3\alpha}k_0^{-3}
\hat P({k a^{\alpha}/ k_0})\ \, ; \, 
\xi(x,\tau)=\hat\xi({x/x_0a^\alpha})\, ,
\label{psim}\end{equation} 
where $\alpha=2/(3+n)$;
$k_0,\, x_0$ are constants which must be determined from the initial 
conditions; and $\hat P, \, \hat \xi$ are unspecified dimensionless
functions. It is easy to verify that the linear spectrum 
$P_{\rm lin}(k,\tau)\propto a^2 k^n$ is consistent with the functional
form of equation (\ref{psim}). This functional form provides a
strong constraint on the evolution of the spectrum, as $P$ is no 
longer an arbitrary function of two independent variables $k$ and
$a$, but only of the variable $ka^\alpha$. 

The results for the scaling of the power spectrum are shown in Figures
4 and 5. The self-similar scaling 
given by equation (\ref{psim}) with $n=-2$ is compared with the 
velocity scaling. The two scalings are distinguished by the way
in which characteristic wavenumbers $k_c$ scale with time: 
$k_c\propto a^{-\alpha}$, with $\alpha=2$ for self-similar scaling,
and $\alpha=1$ for the velocity scaling, as discussed at the
end of Section 2. In Figure 4 we show the 
measured power spectrum at eight different times corresponding to
a range of $k_{nl}(a)=58-7.25$ (in units of $2\pi/L$). 
Also shown as the dotted curves
is the power spectrum from the fourth from last time, re-scaled 
self-similarly in the left panel, and according to the velocity
scaling in the right panel. The results show that the 
measured spectra are consistent with self-similar scaling and show 
departures from the velocity scaling which causes the spectrum to
grow too slowly at high-$k$. 
It should be noted that there are apparent 
departures from self-similar scaling at low-$k$. We believe these 
arise due to the statistical fluctuations resulting from the
small number of modes present in the low-$k$ bins. These can be
seen even in figure 1 where the power at the lowest $k$ does not
agree with the linear spectrum. At the latest times however this
discrepancy extends to $k\simeq 10$. A possible explanation is
the missing power from long-wave modes, which, as one might expect
from the action of nonlinear mode coupling, appears to affect
the spectrum at increasing $k$ at later time outputs.
However this issue merits further exploration as it contradicts
our own estimate of $k>4$ being an adequate cut-off at the low-$k$ end. 

Figure 5 shows the same feature in a
slightly different way: the power spectrum curves at the eight
different times are re-scaled so that they would lie along a single
curve if they followed the predicted scaling. Again, the left panel
which uses self-similar scaling shows that the curves almost
coincide. The right panel showing the velocity scaling has significant
differences between the different curves even at intermediate
$k$ on which the spectrum is measured very accurately. 
In figures 4 and 5 the agreement with the self-similarly
scaled spectra is typically within $20\%$, and better than
$40\%$ over nearly the entire range (excluding only 
the lowest $k$ where fluctuations occur due to the small number of modes).
For the velocity scaling, the agreement is typically at the $50\%$
level and is about a factor of 2 if one uses the earliest and latest
output times. The curve traced 
out by the scaled N-body spectra in the left-panel of figure 5 
give the empirical $\hat P$ of equation (\ref{psim}). 

To highlight the numerical difficulty of
demonstrating self-similar scaling for the $n=-2$ spectrum, 
in Figure 6 we show the results from
the smaller of our P$^3$M simulations, labeled B in Table 1. 
The figure shows the scaling of the spectrum in exactly the same
way as Figure 5. On comparing the two panels in 
Figure 6 it is evident that the dynamic range of simulation B is
not adequate to discriminate between the self-similar and velocity
scalings. The scatter in the different curves due to numerical 
effects at low and high-$k$ (which overlap with intermediate-$k$
spectra at other times because of the re-scaling made in the plots)
is comparable to the difference in the self-similar and velocity
scalings. It is also clear that without comparing the two scalings,
it is very difficult to distinguish the effects of limited dynamic range 
from real departures from the underlying scaling. 

The resolution of simulation B shown in Figure 6
is comparable to the simulations used by Lacey \& Cole
(1994) and Jain, Mo \& White (1995). Lacey \& Cole used a
P$^3$M simulation with $128^3$ particles and $\epsilon=L/4160$, 
while the simulation used by Jain, Mo \& White was simulation B
itself. Colombi, Bouchet \& Hernquist (1995) used 
a treecode simulation with $64^3$ particles and a spatial resolution
somewhat larger than that of simulation B. 
The resolution limitations of the PM simulation C are even more
severe than those of simulation B. At late times the power at high-$k$
is significantly smaller than that required for self-similarity. As 
discussed at the end of the previous section, Figure 3 
shows that for simulation C the power on $k>80$, i.e. on $k$ larger 
than $1/6$th the Nyquist frequency of the PM grid, is artificially 
suppressed by $50\%$ at the latest $a$ with $k_{nl}(a)=7.25$ (in units of
$2\pi/L$). The $n=-2$ simulation of Padmanabhan et al. (1995) is a PM
simulation with $240^3$ particles on a $720^3$ PM mesh, and used a 
staggered mesh scheme. Thus its spatial resolution is about twice
superior to that
of simulation C, but is still considerably
worse than that of simulation B (since its PM grid spacing is 
$>2 \epsilon$ for simulation B) which is shown in Figure 6
to lack the resolution to test self-similar scaling. 

\subsection{Shape of the Nonlinear Spectrum}

Having shown that simulation A for the $n=-2$ spectrum 
follows self-similar scaling, we can characterize the spectrum 
at any time by the self-similar functional form of equation (\ref{psim}).
The qualitative features of this shape can be seen in Figure 1
which shows the nonlinear and linear $P(a,k)$ at five different times,
and in Figure 5 which shows 
the self-similar shape of the spectrum traced out by the measured
spectra at eight different times. At sufficiently low $(k/k_{nl})$ the 
spectrum follows the linear shape $\propto k^{-2}$. At $k\sim
k_{nl}$ the spectrum rises above the linear spectrum and then at
still higher $k$ it again approaches the shape $\propto k^{-2}$. 
The enhancement of the nonlinear spectrum relative to the linear
spectrum is a factor of $\simeq 1.5$ at $k=k_{nl}$,  and 
approaches a factor $\simeq 5$ at $k\gsim 10 \, k_{nl}$. It maintains
this enhancement up to the highest $k$ measured. 

The shape of $P(k)$ measured at $k\gsim 10\, k_{nl}$ is in agreement with the 
prediction of the stable clustering ansatz. This ansatz relies on the
assumption that the mean pair velocity in physical coordinates is
zero. As shown by Davis \& Peebles (1977) this leads to the prediction
that the asymptotic form of $P(k)$ at high-$k$ is: $P(a,k)\propto
a^{6/(5+n)} k^{-6/(5+n)}$. For $n=-2$ this is: $P\propto a^2 k^{-2}$,
in excellent agreement with the measured slope shown in Figure 5. Note
that once self-similarity is taken to be valid, the 
growth in $a$ is fixed by the $k^{-2}$ shape. 
Note also that for $n=-2$ the shape of the spectrum in the stable 
clustering regime is the same as that of the linear spectrum. 

The transition from the linear regime to the deeply nonlinear regime of
stable clustering is poorly understood analytically, since
perturbation theory breaks down at $k\sim k_{nl}$. However a 
semi-empirical prescription for obtaining the nonlinear $\xi$ 
for arbitrary initial spectra
was proposed by Hamilton et al. (1991), and extended to the
power spectrum by Peacock \& Dodds (1994). Jain, Mo \&
White (1995) and Peacock \& Dodds (1996) have refined the 
prescription to take into account a dependence on the initial spectrum. 
They provide fitting formulae
for the dimensionless nonlinear spectrum at a given time,
$\Delta(a,k)=4\pi k^3 P(a,k)$ in terms of the linear spectrum 
$\Delta_L(a,k_L)=4\pi k^3P_{\rm lin}(a,k_L)$ evaluated at a
smaller wavenumber $k_L$ as indicated.  The two wavenumbers
$k_L$ and $k$ are related by $k=[1+\Delta(k)]^{1/3}k_L$. 
One has to first obtain $\Delta(k)$ for given $k_L$ and $\Delta_L$
using equation \ref{Delta} below, and then substitute it into the
relation above to get the value of $k$.
The relation between $\Delta(a,k)$ and $\Delta_L(a,k_L)$ is 
(Jain, Mo \& White 1995):
\begin{equation}
{\Delta (a,k)\over B(n)}= 
\Phi \left\lbrack {\Delta_L (a,k_L)\over B(n)}\right\rbrack ,
\label{Delta}\end{equation}
where the constant $B(n)=[(3+n)/3]^{1.3}$, and the function $\Phi$ is
\begin{equation}
\Phi (x)= x\left( {1+0.6 x+x^2 -0.2 x^3 -1.5 x^{7/2} + x^4 \over
      1+0.0037 x^3}\right)^{1/2} .
\label{Phi}\end{equation}
The formulae proposed by Peacock \& Dodds (1996) give similar
results as the equations above, though the nonlinear asymptote
is somewhat lower for the $n=-2$ spectrum. 

The fitting formula given above can be tested by using the nonlinear
spectrum measured in our simulations. The results for the spectrum
from simulation A are shown in Figure 7. The dimensionless spectrum
$\Delta(a,k)$ is plotted against $\Delta_L(a,k_L)$ for the 
output times shown in Figure 1. The solid curve shows the fit
provided by equations (\ref{Delta}) and (\ref{Phi}). The 
dashed curve is obtained from the fitting formula of Peacock \& Dodds (1996).
The dotted
lines show the slope of the linear and stable clustering predictions. 
The agreement between the N-body spectra and the solid curve is
adequate, and is better than about $ 30\%$ over almost the entire
range plotted (note that the accuracy of the N-body spectrum at the
smallest and largest amplitudes shown is about $20\%$). 
The N-body spectrum is enhanced relative to the fit at large
amplitudes. This is in the same sense as the discrepancy noted by 
Jain, Mo \& White (1995) who observed that the $n=-2$ spectrum
has a slightly steeper nonlinear shape in the
$\Delta(k)-\Delta_L(k_L)$ plot as compared to spectra with larger 
$n$. The results also show that the N-body spectra approach the stable 
clustering slope at amplitudes of $\Delta(k)\gsim 50$. 

%(???) Figure IV shows the scaling of the autocorrelation
%function $\xi(x,\tau)$ for $n=-2$. The measurement of $\xi$ is more 
%accurate at small scales, as the interpolating grid required for
%computing the power spectrum is not needed. The limitation is only
%the distance at which force softening becomes significant, 
%about $2 \epsilon$. Hence it is useful to
%check that the scaling inferred from the power spectrum
%is verified by $\xi$. 

\section{Alternate Measures of Scaling}

The scaling properties of the matter distribution for scale-free
conditions can be measured from any appropriate statistical 
measure. The power spectrum is the second moment of the density field,
and is therefore one such statistic. The previous section demonstrated
that it is only with very large simulation that the scaling of the
power spectrum can be accurately tested. Higher order statistics are 
even harder to measure accurately. Of course every statistical measure
which relates to the growth of perturbations should measure the same
scaling properties, but to measure them in N-body simulations for 
``difficult'' spectra like $n=-2$, it is important to check more than
one statistic. In this section we measure
the scaling properties of the full density field by using 
a different approach. Rather that compute a statistic and then measure
its scaling in time, we compute the fractional departure of the phase and
amplitude from linear evolution
mode by mode. The degree of nonlinearity thus measured is used to
define characteristic nonlinear
scales for the phase and amplitude. Thus by using both the phase and
amplitude, and by using the nonlinearity of individual modes rather
than an averaged quantity like the power spectrum, we are able to 
probe the scaling of the full density field more completely.

We use the data for $\hat \delta(\vec k,a)$ to compute the 
amplitude $\Delta$ and phase $\phi$ given by,
\begin{equation}
\hat \delta(\vec k,a)=\Delta(\vec k,a)e^{i\phi(\vec k,a)}. 
\label{sim1}\end{equation}
$\Delta$ and $\phi$ are the basic variables used for the scaling
analysis in this section. In Figures 
%\ref{fig:1} to \ref{fig:3} 
8-10 the trajectories
of the phase and amplitude of individual Fourier modes as a function of
time are shown for the $n=-2$ simulation. 
Each of these figures has four panels, and each panel shows the evolution
of five modes, chosen so that they represent a large range in $k$. 
These figures give an idea of how individual modes evolve, in contrast to 
the regular behavior shown by the statistics computed from them. 
Even at relatively early times when  most statistics obey linear 
behavior the amplitudes and phases can be seen to follow quite jagged
paths, with the amplitude even showing negative growth for some
time intervals. This suggests that there is more information to be
mined than is provided by conventional statistics. 

\subsection{Measuring the Phase Shift}

A technical problem in the measurement of
the phase arises because conventionally the phases are defined modulo 
$(2 \pi)$. The phase trajectories that result are
shown in Figure 8,
%\ref{fig:1}
where all the phases lie between $-\pi$ and
$\pi$. This is how the phases from N-body data have been 
computed in previous studies (Ryden \& Gramann 1991; Scherrer, Melott
\& Shandarin 1991; 
Suginohara \& Suto 1991). However, following the phase trajectories in 
Figure 8
%\ref{fig:1} 
makes it clear that this will have the effect of randomizing
the phases at late times  relative to their initial values, even if
the actual growth were monotonic. This is because even a small change 
in the phase, 
$\delta\phi$, could cause it to be mapped to a value indicating a change of 
$(2\pi-\vert \delta\phi\vert)$. 

If the trajectories could be obtained with arbitrarily small increments in 
$a$ then such artificial mappings could be un-done, 
and the phases plotted without constraining them between 
$-\pi$ and $\pi$. 
Since outputs are available only at discrete values of $a$, there is a 
two-fold ambiguity in defining the 
phase. Consider the phase values at two successive $a$'s for a given $\vec k$:
$\phi(\vec k,a_j)$ and $\phi(\vec k,a_{j+1})$, defined in the usual way to 
lie between $-\pi$ and $\pi$. Let $\delta\phi_j=\phi(\vec k,a_{j+1})-
\phi(\vec k,a_j)$. 
An alternate value for the phase at $a_{j+1}$ is
$\phi'(\vec k,a_{j+1})=\phi(\vec k,a_j) \pm (2\pi-\vert\delta\phi_j\vert),$
where the sign in positive if $\delta\phi_j<0$ and vice versa.
To choose between $\phi$ and $\phi'$, we
follow the trajectory of each mode, and at each successive value of $a$, we
define the phase by taking the magnitude of the change in phase to be 
the smaller of $\vert\delta\phi_j\vert$ and $(2\pi-\vert\delta\phi_j\vert)$. 
The result is shown in Figure 9.
%\ref{fig:2}
As long as the
typical changes in phase at successive times are less than $\pi$, this 
procedure is a reasonable 
way of extending the range of $\phi$. As we shall see, this 
considerably extends the degree of phase nonlinearity accessible to our 
analysis. 

\subsection{Scaling of the Phase and Amplitude}

Following Ryden \& Gramann (1991) and Gramann (1992), we define the following
statistics as a measure of the degree of nonlinear evolution. For the phase
we define the mean deviation from the initial phase, $\overline{\delta\phi}
(k,a) \equiv \langle\vert \delta\phi
(\vec k,a)\vert\rangle=\langle\vert
\phi(\vec k,a)-\phi(\vec k,a_i)\vert\rangle$, where
$a_i$ is the initial value of $a$. The averages
indicated are performed over the different modes within a shell in $k-$space
whose wavenumbers lie between $(k-0.5)$ and $(k+0.5)$. 
For the amplitude we simply measure the mean amplitude $\langle\Delta(\vec k,a)
\rangle$ within each shell in $k-$space. 
%The results are plotted in 
%Figures 4 and 5
%%\ref{fig:4} and \ref{fig:5} 
%for $n=-2$, and Figures 11 and 12
%%\ref{fig:11} and \ref{fig:12} 
%for $n=0$. These results are interesting in themselves, 
%but here we limit ourselves to using them to define
%characteristic wavenumber scales, which are then used to test for 
%self-similar scaling. For Figures 4 and 11,
%%\ref{fig:4} and \ref{fig:11}
If we had
used the phase trajectories as shown in Figure 8
%\ref{fig:1} 
(i.e., defined 
to lie between $-\pi$ and $\pi$), then at late
times $\overline{\delta\phi}(k,a)$ would have reached
a maximum value $2\pi/3$ ---
this corresponds to a distribution of $\phi(\vec k,a)$ that is uncorrelated 
with $\phi(\vec k,a_i)$. 
However, we find
%\ref{fig:4} 
that $\overline{\delta\phi}$ shows systematic growth well beyond $2\pi/3$. 
Thus with the phase information that we have generated, previously 
unexplored aspects of phase evolution in the deeply nonlinear regime can
be addressed. 

To analyze self-similar scaling, we define two characteristic wavenumbers. 
The first, denoted $k_c(a,\phi_c)$,
is defined by setting $\overline{\delta\phi}(k,a)=\phi_c$, where $\phi_c$ 
is a constant. The second, denoted $k_c(a,\Delta_c)$ is defined using the
amplitude as follows: 
\begin{equation}
\left\langle{\vert\Delta(\vec k,a)-\Delta_1(\vec k,a)\vert
\over\Delta_1(\vec k,a)}\right\rangle = \Delta_c\, ,
\label{sim2}\end{equation}
where $\Delta_1(\vec k,a)$ is the linear solution for $\Delta$, and
$\Delta_c$ is a dimensionless constant.
Thus $k_c(a,\Delta_c)$ is the wavenumber at which the fractional departure
of the amplitude from the linear solution is $\Delta_c$. 
These statistics involve summing the magnitudes of the departures from 
linear behavior
for each mode within a given $k-$shell. Hence they probe the degree of 
nonlinearity more directly than if a statistic was computed first, and then 
its departure from the linear solution was calculated. 

The analytical prediction for the variance of the phase shift 
given in Paper I is
\begin{equation}
\left\langle\left[\delta \phi(\vec k,\tau)\right]^2
\right\rangle= {4 \pi\over 3}
a(\tau)^2\, k^2\, \int\! d k_1 \, {P_{\rm lin}(k_1)} \, .
\label{ph10}\end{equation}%4.30
%For later reference we also give the result for the mean magnitude of the
%phase shift, 
%begin{equation}
%\langle\vert\phi(\vec k,\tau)=
Thus the leading order solution for 
$\hat \delta(\vec k)$ involves a growing (and, for $n<-1$, divergent)
phase shift, but there are no contributions to the amplitude at this
order. With some further assumptions we were able to show analytically
that the amplitude does not diverge provided $n>-3$, and should
therefore show the standard self-similar scaling. We did not however
obtain an analytical expression analogous to (\ref{ph10}) for its growth. 

Figure 11
%\ref{fig:6} 
is a plot of ${\rm log}[k_c(a,\phi_c)]$  vs. $a$  with $n=-2$,
for $4$ different values of $\phi_c$. Also shown in the plot are the scalings
expected from self-similarity, $k 
\propto a^{-2}$, and the scaling resulting from
the solution for $\delta\phi$ given by equation (\ref{ph10}), 
$k\propto a^{-1}$. This is the same as the velocity scaling referred
to in Sections 2 and 3. 
The plots show that for $\phi_c=\pi$ and $\pi/2,$ the  velocity scaling is 
closely followed, typically to better than $20\%$; but for the lower 
values $\phi_c=\pi/4$ and $\pi/8$, self-similar
scaling is more closely followed at the same level of accuracy. We verified 
that the trends did reflect
a gradual transition in the scaling of $k_c(a,\phi_c)$ by plotting a larger
range of $\phi_c$ down to $\phi_c=\pi/20$. 

The dual scaling behavior shown by the phase shift can be interpreted as
follows. For a given $k$, at early times as the phase just begins to depart
from the linear solution (in which $\phi$ remains constant in time), 
its evolution is dominated by
perturbative or other weakly nonlinear effects. These effects in general 
involve the coupling of a range of values of $k'$, mostly in the vicinity of 
$k'=k$, and obey the standard self-similar scaling. However at late times 
the phase shift is dominated by the bulk flow due to 
the longest waves in the box. The resulting 
phase change is given by equation (\ref{ph10}) --- it is a smooth function of
$a$ and $k$, and therefore dominates the more stochastic, dynamical
components of the phase change at late times. 
%We have verified that the 
%magnitude of $k_c(a, \phi_c)$ for $\phi_c=\pi/2$ is in good agreement with
%the prediction from equation \ref{ph10}. 
Thus the phase shows behavior that we can interpret as arising from a 
combination of the kinematical divergence, and a dynamical,
nonlinear component which obeys the standard self-similar scaling. 
The former drives $k_c(a,\phi_c)$ to the $k_c\propto a^{-1}$ scaling
at late times, and the latter to the $k_c\propto a^{-2}$ scaling at
early times. 

Figure 13 shows $k_c(a,\phi_c)$ vs. $a$ for the $n=0$ spectrum. 
These results are from a P$^3$M simulation performed by S. White, 
with $100^3$ particles and $\epsilon=L/2500$. The $n=0$ spectrum 
%\ref{fig:13} 
shows only one behavior, the self-similar scaling, $k\propto a^{-2/3}$.
This is expected as the linear bulk velocity does not diverge, therefore the 
longest waves in the box do not dominate the phase shift at any time. 

For the amplitude scaling, Figure 12
%\ref{fig:7} 
shows a plot of ${\rm log}
[k_c(a,\Delta_c)]$ vs. $a$ with $n=-2$, for $\Delta_c=0.25,0.5,1,2$. 
For sufficiently high $k$, all four curves closely follow the standard
self-similar scaling, $k \propto a^{-2}$, to better than $20\%$ accuracy. 
This is consistent with the 
results for the power spectrum, as one would expect since the 
power spectrum measures the variance of the amplitude. 
All the curves show a departure from the $k \propto a^{-2}$
scaling at low $k$, for $k<10$, with the discrepancy exceeding
a factor of 2 in the amplitude. This most likely indicates that
the absence of power on modes with wavelengths
larger than the box-size has slowed the growth of modes which would otherwise
be enhanced by coupling to modes longer than the box. 
Thus the standard self-similar solution for
$n=-2$ is obtained only on scales significantly smaller than the box-size. 
For $n=0$ the self-similar scaling $k\propto a^{-2/3}$ is again shown 
convincingly in Figure 14. 
%This would explain why the scaling results for $n=-2$ in three 
%dimensions obtained previously using smaller simulations were not very 
%convincing. 

Our results are in partial disagreement with those of Gramann (1992). 
She found that for $n=-1$ in two dimensions (the analog of $n=-2$ in 
three dimensions), the standard self-similar scaling is broken for both
the phase and amplitude. Our results for the phase scaling are consistent with
hers, but the amplitude scalings are quite different: our results show good
agreement with the scaling $k_c\propto a^{-2}$, whereas hers agree with 
$k_c\propto a^{-1}$. 
Since the statistics that we have measured are exactly the same as hers,
it is difficult to explain
the origin of the disagreement. It is conceivable that there are basic
differences in the dynamics in two and three dimensions, but this is not
reflected in the analytical results of Paper I. It is also possible that
effects of the finite box-size are more prominent in two dimensions.

\section{Conclusion}

In Section 2 we summarized the motivation for 
examining the self-similar scaling of scale free spectra, $P(k)\propto k^n$,
for $-3<n<-1$. 
We have examined this issue through analytical techniques in Paper I
and N-body techniques in this paper. In Paper I we found 
through two different approaches that the self-similar scaling of
the amplitude of the density, and therefore of any measure of
dynamical evolution, is preserved for $-3<n<1$. 

In this paper we have tested the scaling of $n=-2$ scale 
free simulations.
In Section 3 we discuss the issues of numerical resolution
involved in such tests by comparing the resolution of three
different simulations. We find that for the $n=-2$ spectrum
we need the full resolution of our largest simulation, a P$^3$M
simulation with $256^3$ particles to measure its scaling properties. 
Section 4 shows that self-similar
scaling is verified by the power spectrum measured from this 
simulation. Figures 4 and 5 demonstrate the existence of this
scaling, and contrast it with an alternate scaling driven by
the large-scale velocity field. We have compared our results with
the recent work of Colombi, Bouchet \& Hernquist (1995);
Jain, Mo \& White (1995); Lacey \& Cole (1994); and 
Padmanabhan et al. (1995). We find that the shape of the nonlinear
spectrum is well fit by the formulae proposed by Jain, Mo \& White
(1995) and Peacock \& Dodds (1996), though the formulae slightly
underestimate the spectrum in the strongly nonlinear regime. 
The shape of $P(k)$ at high-$k$ agrees with the stable
clustering prediction $P\propto k^{-2}$. 

The second test of the scaling properties of the density field 
made in Section 5
relies on the mode-by-mode evolution of the amplitude and phase. 
At sufficiently late times the phase shift obeys the velocity scaling,
consistent with the solution found in Paper I. At early times the 
evolution of the phase shift is consistent with the standard self-similar
scaling. Thus the phase shift arises from a combination of kinematical effects 
due to large scale flows which dominate at late times, 
and genuine dynamical effects which dominate at relatively early times 
--- in the weakly nonlinear stage of its evolution. 
The scaling of the amplitude follows the standard self-similar 
form, except at wavenumbers
$k\lsim10$ (in units of $2\pi/L$) which we believe is a numerical 
limitation due to the finite size of the box. This is consistent
with the analytical results of Paper I and with the results for the 
power spectrum. 

In combination with the analytical analysis of Paper I, our results
lead us to conclude with 
some confidence that the self-similar evolution of the density contrast 
is preserved for $-3<n<-1$. 
The kinematical interpretation for the scaling of the phase
shift provides a useful guide to identifying statistics susceptible to such
effects. The rms displacement of particle positions is an example of a 
statistic which would be dominated by the bulk motions from long wave modes
for $n<-1$, and must therefore be used with caution as a measure of 
nonlinear evolution. 

\acknowledgements

We thank Simon White for providing the results of his 
scale free simulations, and for several useful suggestions. 
We also acknowledge useful discussions with Shep Doeleman, Mirt Gramann,  
Alan Guth, Ofer Lahav, Adi Nusser, Bepi Tormen and, 
especially, David Weinberg. This work was supported by NSF grant
AST-9529154 and a grant of supercomputer time from the National
Center for Supercomputing Applications. BJ acknowledges support
from NASA through the LTSA grant NAG 5-3503.
%Supercomputing time was provided by the Cornell National Supercomputer 
%Facility and the National Center for Supercomputing Applications.
%This work was supported by NSF grant AST90-01762.

%\clearpage
%\centerline{\bf Figure Captions}

\begin{figure}[p]
\vspace*{11.3 cm}
\caption{
%\noindent Fig. 1:
The power spectrum $P(k)$ vs. $k$ measured from simulation A 
at five different times. The solid curves show the N-body power 
spectrum for $k_{nl}(a)=58, 29,
14.5, 7.25$. The wavenumber $k$ is in units 
of $2\pi/L$. The dotted lines show the 
linear spectrum $P\propto a^2 k^{-2}$ at the same times. The 
N-body spectrum at low-$k$ shows statistical fluctuations due to the 
small number of modes available to measure it. At high-$k$ the 
spectrum is plotted up to $1/2$ the Nyquist frequency of the 
FFT grid, $k=512$. 
}
\includegraphics{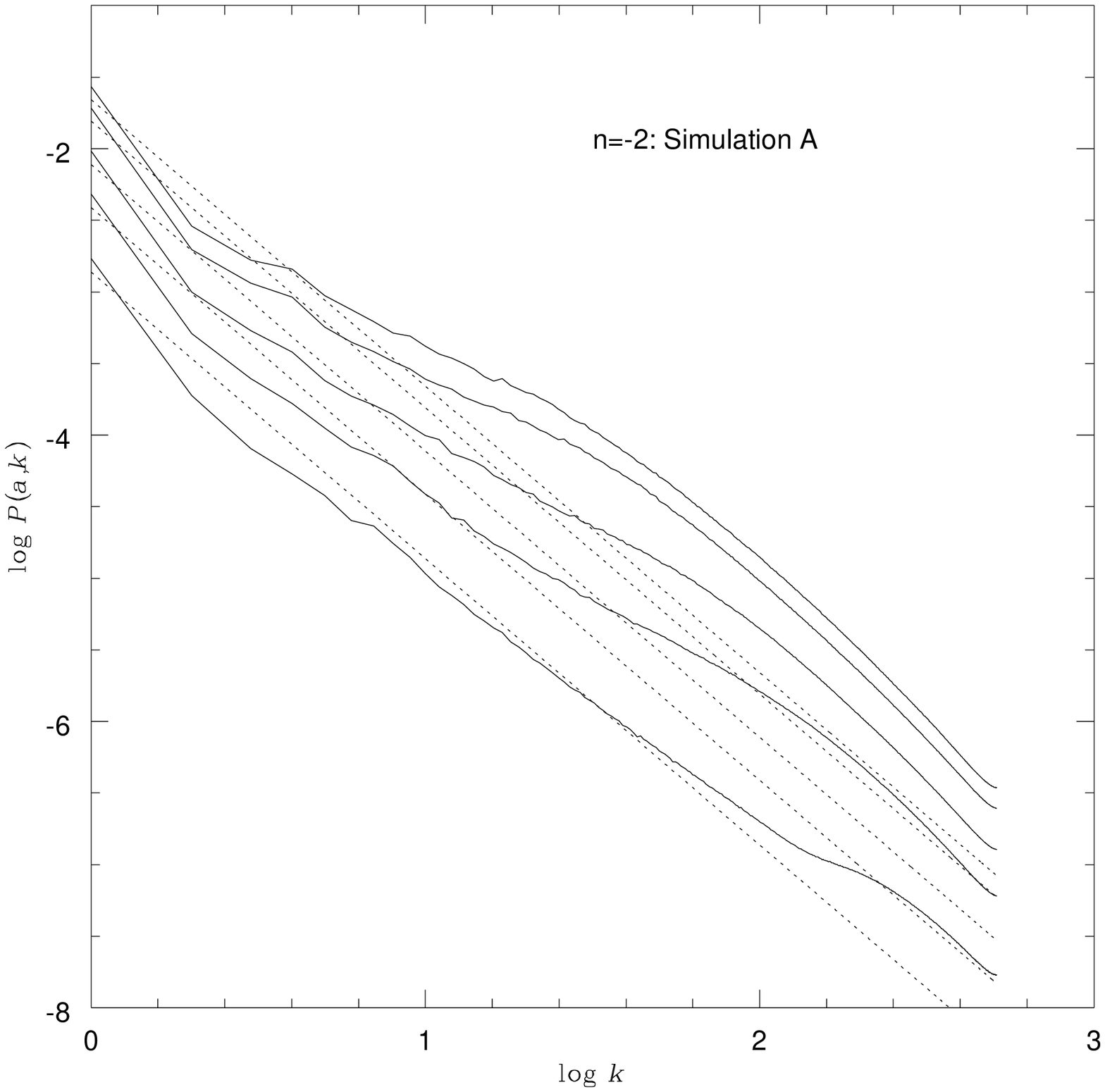}
\label{fig:1}
\end{figure}

\begin{figure}[p]
\vspace*{11.3 cm}
\caption{
%\noindent Fig. 2:
The effect of the FFT grid on $P(k)$ at high-$k$. 
The two solid curves show the power spectrum computed from simulation
B with different grid sizes: $512^3$ and $1024^3$. The spectrum
measured with a $512^3$ grid is plotted with a dotted curve between
$k=257-512$, to show the artificial features introduced by the grid
used to interpolate the real space density. 
}
\includegraphics{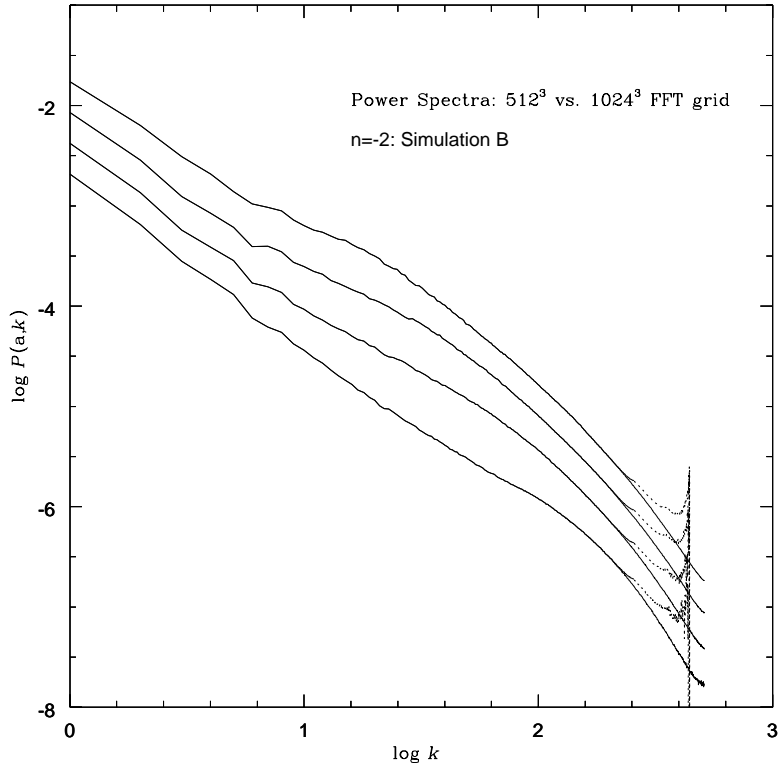}
\label{fig:2}
\end{figure}

\begin{figure}[p]
\vspace*{11.3 cm}
\caption{
%\noindent Fig. 3:
The effect of limited resolution on the power spectrum. 
The solid curves show the power spectrum measured from simulation
C, a PM simulation with the same number of particles as simulation
A, a P$^3$M simulation. The spectra are computed on a $512^3$ grid and
are shown at the same output times as those of Figure 1. The
dotted curves show the spectra from simulation A, computed using a 
$1024^3$ grid. Both spectra are plotted up to $1/2$ the Nyquist 
frequency of the FFT grid. A comparison of the two sets of spectra
shows that as time evolves, the power in the PM simulation is
suppressed on an increasingly large range of 
$k$. 
}
\includegraphics{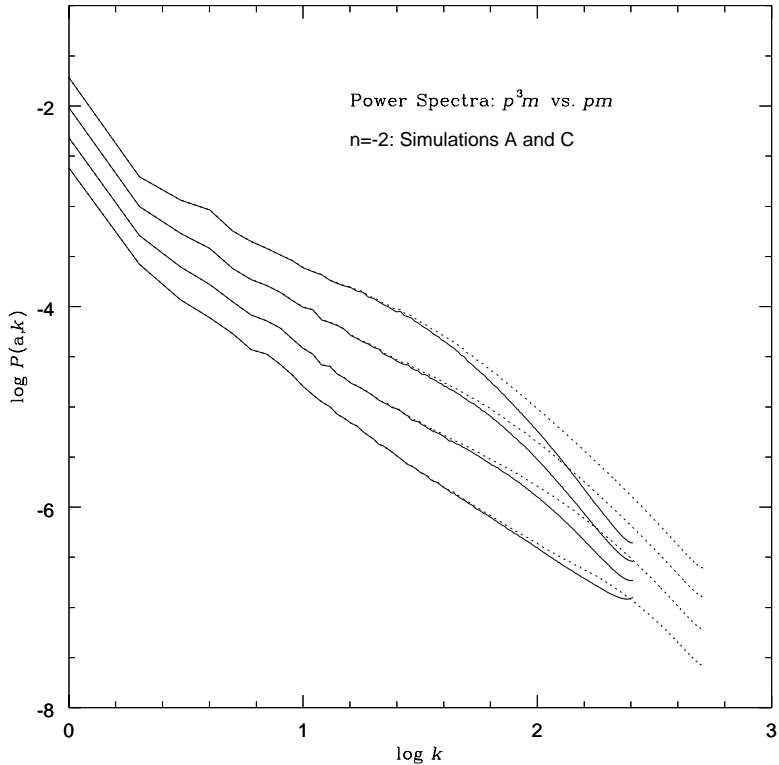}
\label{fig:3}
\end{figure}

\begin{figure}[p]
\vspace*{11.3 cm}
\caption{
%\noindent Fig. 4:
The scaling of the power spectrum for the $n=-2$ simulation A. 
Simulation A is a P$^3$M simulation with $256^3$ particles. 
The solid curves show the N-body power spectrum at eight
different output times spanning $k_{nl}(a)=58-7.25$. 
The dotted curves are all obtained by scaling
the spectrum at the fourth from last output time: the wavenumber 
is scaled as $k\propto a(\tau)^{-\alpha}$, and the spectrum is scaled
to maintain consistency with the linear spectrum. The left panel
shows the results for the standard self-similar scaling with
$\alpha=2$. The dotted curves agree very well with the solid curves 
except at the smallest and largest $k$ owing to the finite resolution
of the simulation. The right panel shows the velocity scaling with $\alpha=1$. 
}
\includegraphics{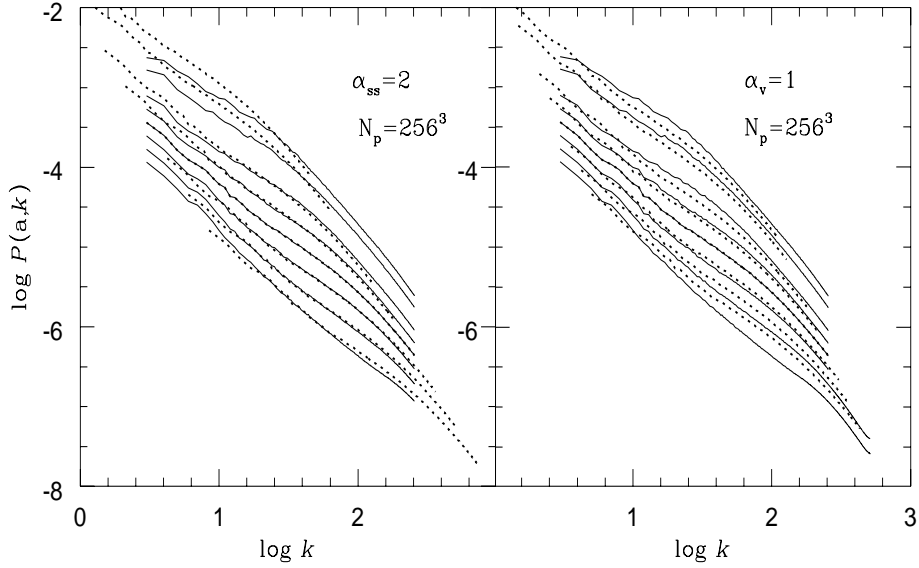}
\label{fig:4}
\end{figure}

\begin{figure}[p]
\vspace*{11.3 cm}
\caption{
%\noindent Fig. 5:
The same power spectra as in Figure 4 are shown, but the scaling is
tested a little differently. The power spectra and wavenumbers at 
the eight output times are scaled so that they would all lie along a
single curve if the scaling were correct. Again the standard
self-similar scaling shown in the left panel works better than the
alternate velocity scaling shown in the right panel. The lower 
dashed line shows the linear spectrum. The upper dashed line has
arbitrary amplitude, but its slope is predicted by the stable
clustering hypothesis. The N-body spectra asymptote to this slope
at high-$k$. 
}
\includegraphics{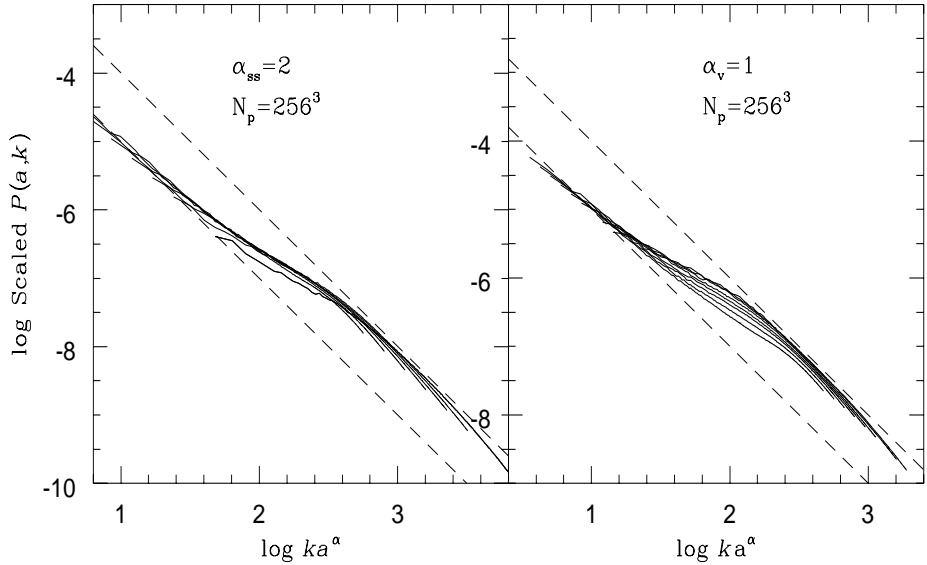}
\label{fig:5}
\end{figure}

\begin{figure}[p]
\vspace*{11.3 cm}
\caption{
%\noindent Fig. 6:
This is identical to Figure 5, but for simulation B 
which has $128^3$ particles and twice as large a spatial resolution
as simulation A. 
The results shown in the two panels are not able to distinguish the 
different scalings. This figure 
demonstrates the need for larger simulations, like the one shown in 
Figures 4 and 5, to effectively test the scaling properties of the 
$n=-2$ spectrum. 
}
\includegraphics{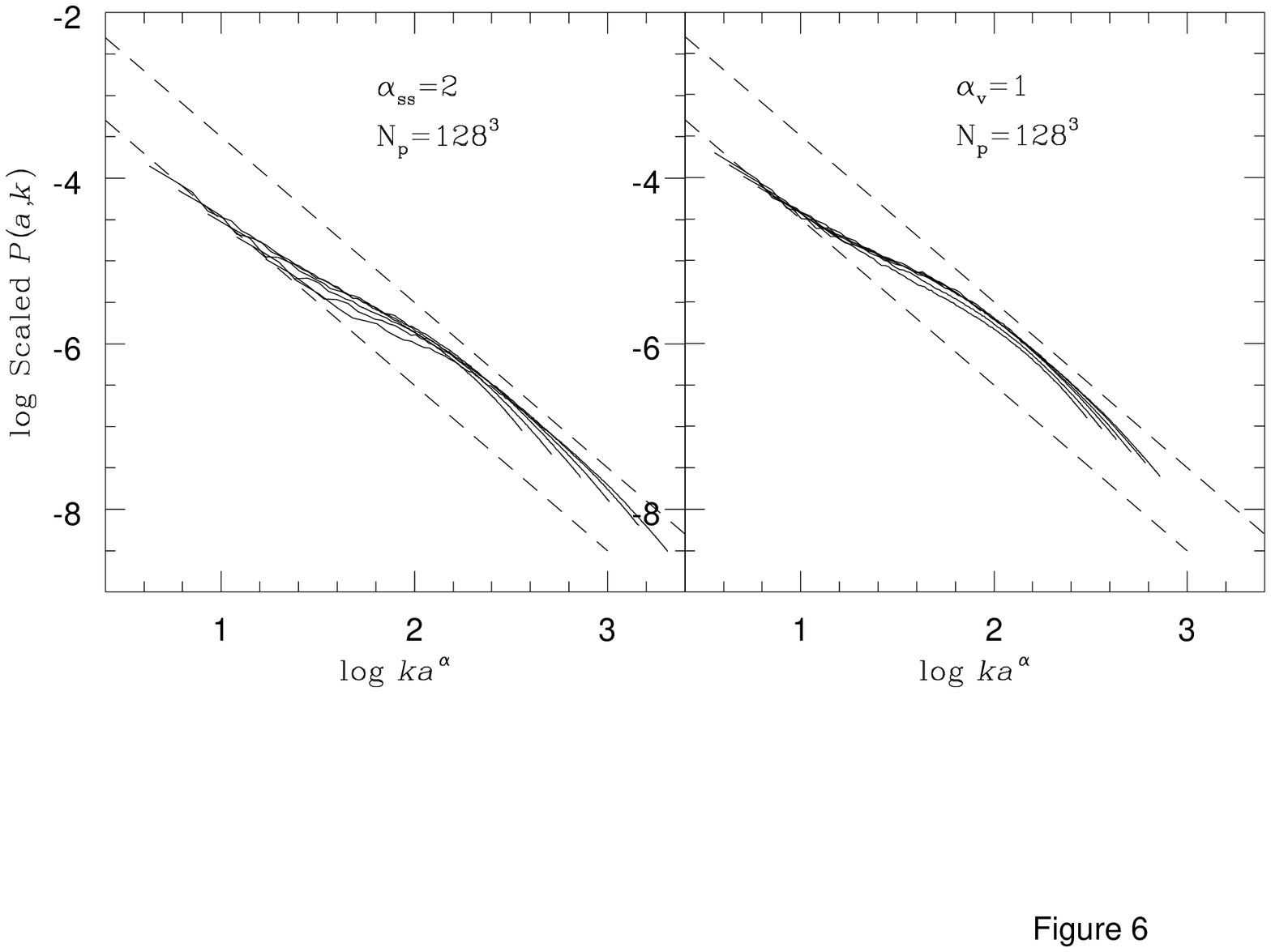}
\label{fig:6}
\end{figure}

\begin{figure}[p]
\vspace*{11.3 cm}
\caption{
%\noindent Fig. 7:
Comparison of N-body spectra with analytical fitting formulae.
The dotted curves show the dimensionless
power $\Delta(k)=4\pi k^3 P(k)$ plotted against the linear power on
wavenumber $k_L$, as described in Section 4.1. The curves are
obtained from the N-body spectrum of simulation A at the same times
as in Figure 1. The solid curve is the fitting formula of Jain, 
Mo \& White (1995) as given in equation
(\ref{Phi}) and the dashed curve is from the fitting formula of 
Peacock \& Dodds (1996). The upper dotted line shows the slope predicted by
the stable clustering hypothesis, while the lower dotted line is the
linear theory relation. 
}
\includegraphics{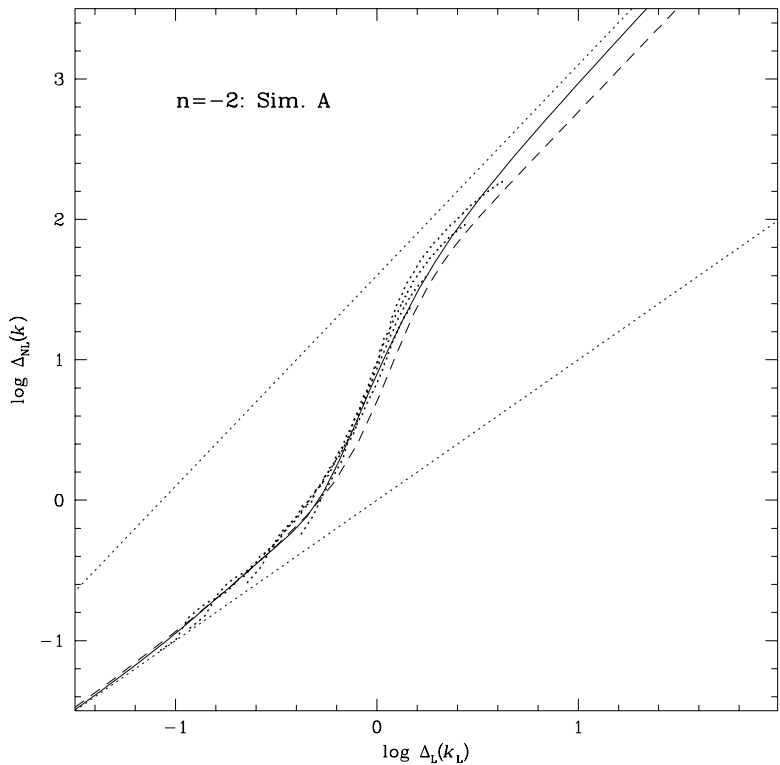}
\label{fig:7}
\end{figure}

\begin{figure}[p]
\vspace*{11.3 cm}
\caption{
%\noindent Fig. 8:
Phase trajectories for the $n=-2$ simulation B. 
The trajectories of the phases $\phi(\vec k,\tau)$
of individual Fourier modes for $n=-2$ are
plotted vs. $a(\tau)$. The magnitudes $k$ of the wavevectors are labeled in 
each panel; the full vectors were chosen as $\vec k = (0, 0, k)$. Within 
each panel $k$ increases in the following order: solid, dotted, dashed, 
long-dashed, dashed-dotted curves. The phase is defined modulo $2\pi$, and 
is therefore constrained to lie between $-\pi$ and $\pi$. 
At $a=0$, the value of $\phi$ at the earliest time has been plotted 
again to show the expected linear behavior. 
}
\includegraphics{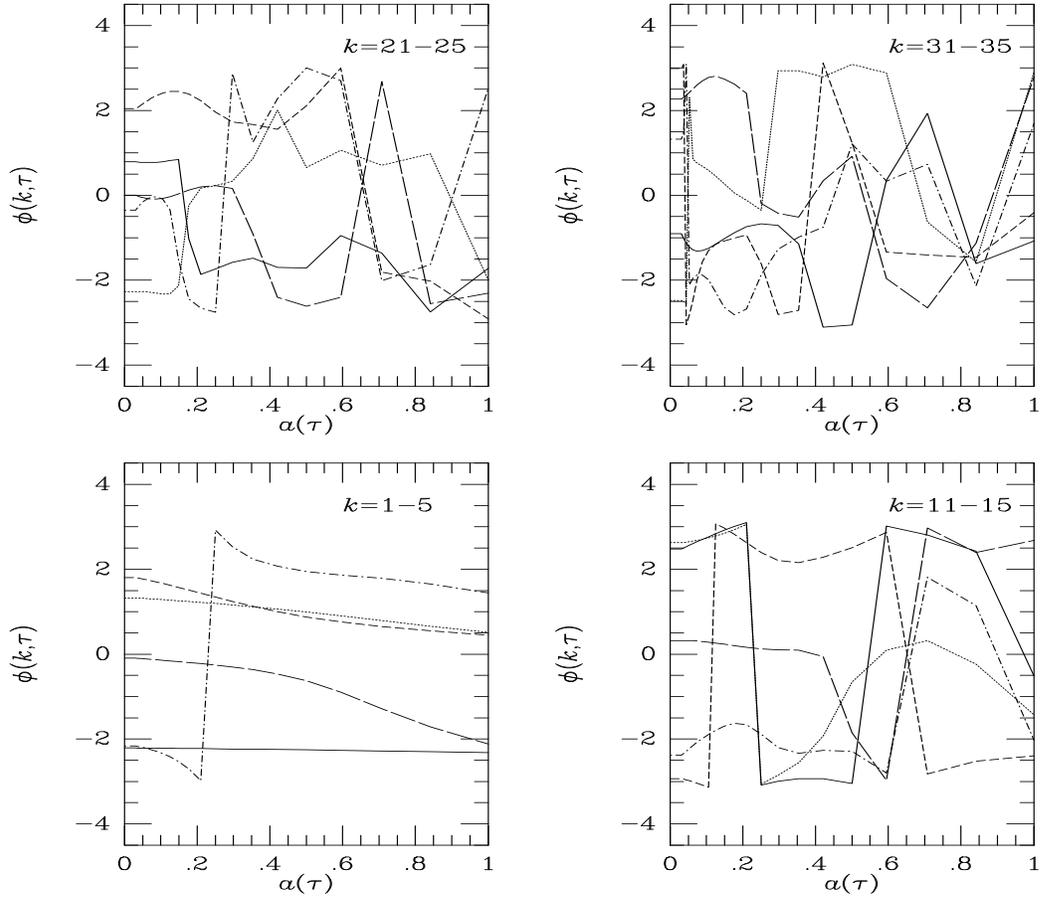}
\label{fig:8}
\end{figure}

\begin{figure}[p]
\vspace*{11.3 cm}
%\noindent Fig. 9:
\caption{
``Re-defined'' phase trajectories for $n=-2$.
The same trajectories as in Figure 8 are plotted, but 
$\phi$ has been re-defined so that it is no longer constrained between 
$-\pi$ and $\pi$ (notice the limits on the y-axis), 
as described in Section 5.1.
In linear theory the phases do not change with time; significant departures
from this can be seen in all but the lowest $k$ modes. 
}
\includegraphics{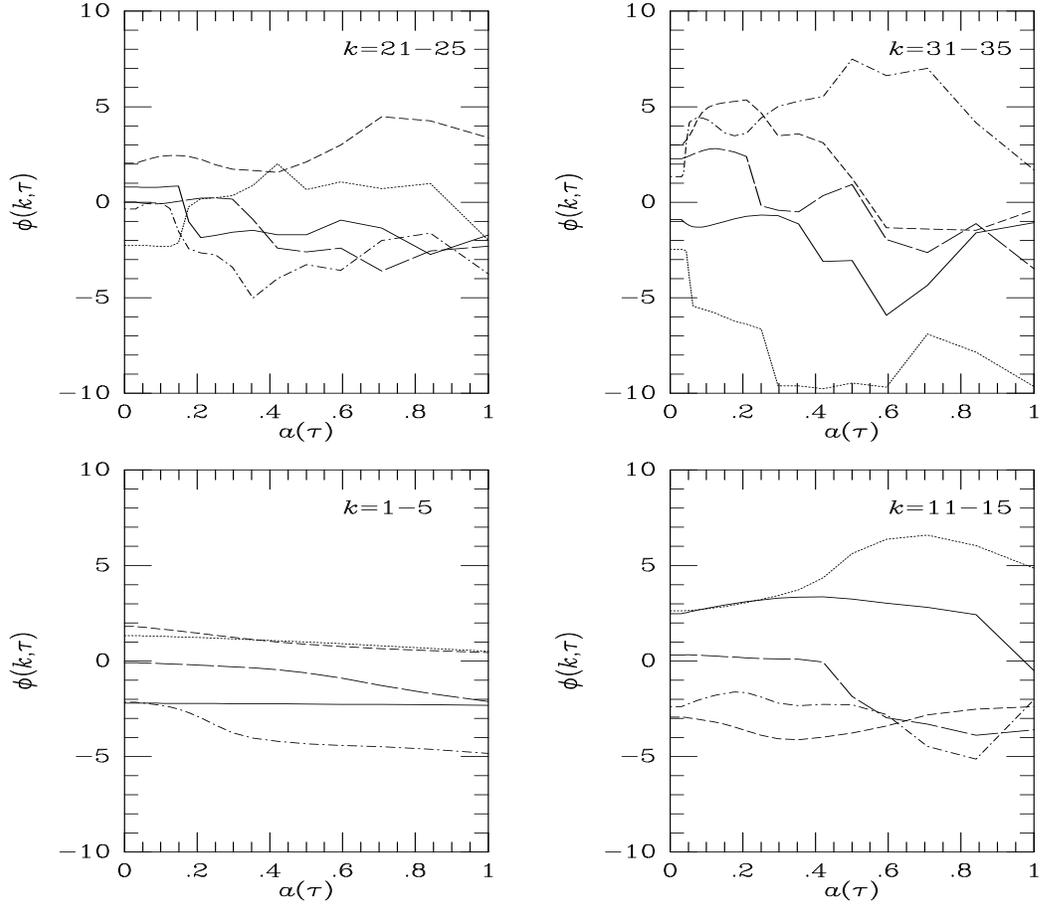}
\label{fig:9}
\end{figure}
%\label{fig:2}
%\end{figure}

\begin{figure}[p]
\vspace*{11.3 cm}
\caption{
%\noindent Fig. 10:
Amplitude trajectories for $n=-2$.
The trajectories of the amplitudes $\Delta(\vec k,\tau)$
of individual Fourier modes for $n=-2$ are plotted vs. $a$, as in Figure 9
%\ref{fig:1} 
for the phase. Note that at early times $\Delta\propto a$:
to check this all the curves have been joined to $\Delta=0$ at $a=0$. 
Therefore the lowest value of $a$ at which departures from a straight line
occur shows nonlinear behavior. 
}
\includegraphics{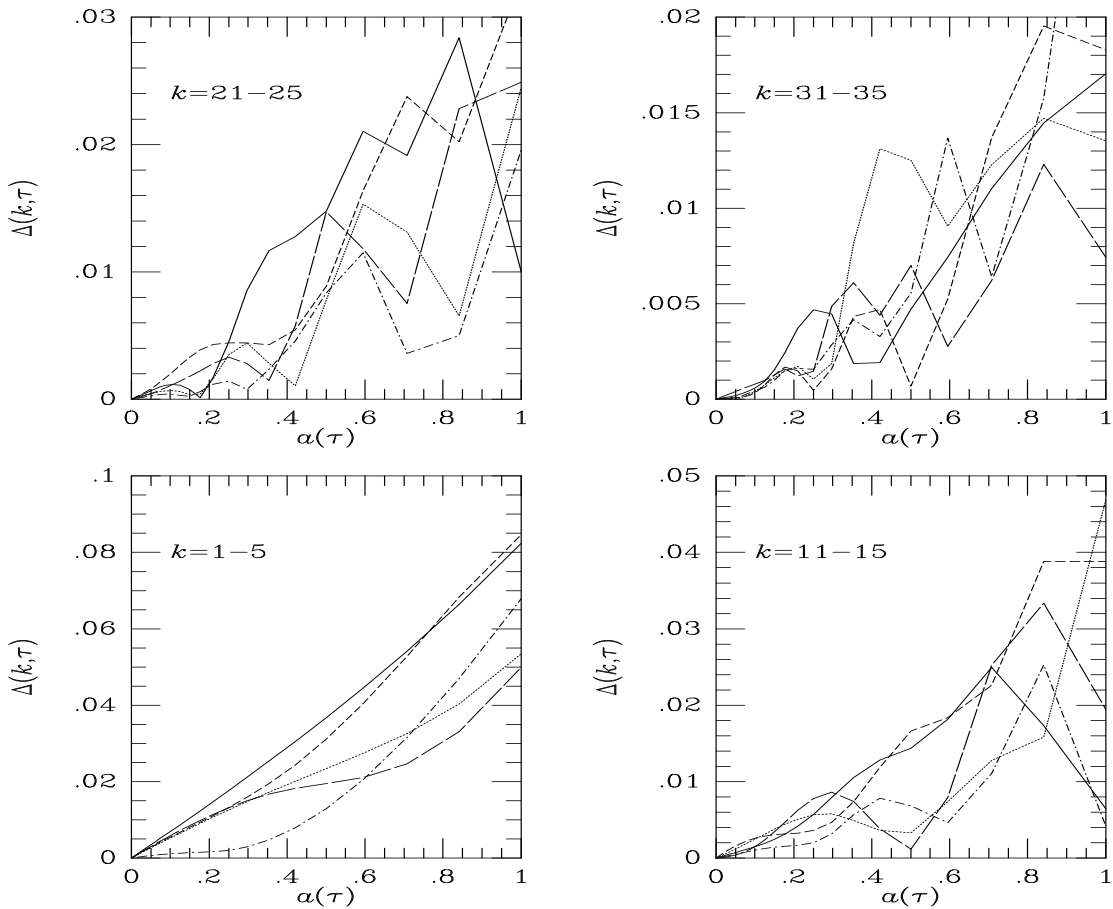}
\label{fig:10}
\end{figure}

\begin{figure}[p]
\vspace*{11.3 cm}
\caption{
%\noindent Fig. 11:
Characteristic scales from the phase for $n=-2$.
The scaling of characteristic wavenumber scales, 
$k_c$ vs. $a$ is shown, as derived from the mean phase deviation.
$k_c(a,\phi_c)$ is the value of
$k$ at which $\langle\vert \delta\phi(\vec k,a)\vert\rangle=\phi_c$. 
The four solid curves correspond to the values of $\phi_c$ labeled
at the top of the plot. Note that for high values of $\phi_c$ (the top $2$
curves), the scaling closely agrees with 
$k_c\propto a^{-1}$, shown by the upper dotted line. 
A transition towards the standard self-similar scaling $k_c\propto a^{-2}$, 
shown by the lower dotted line, occurs for the two lower values of $\phi_c$. 
}
\includegraphics{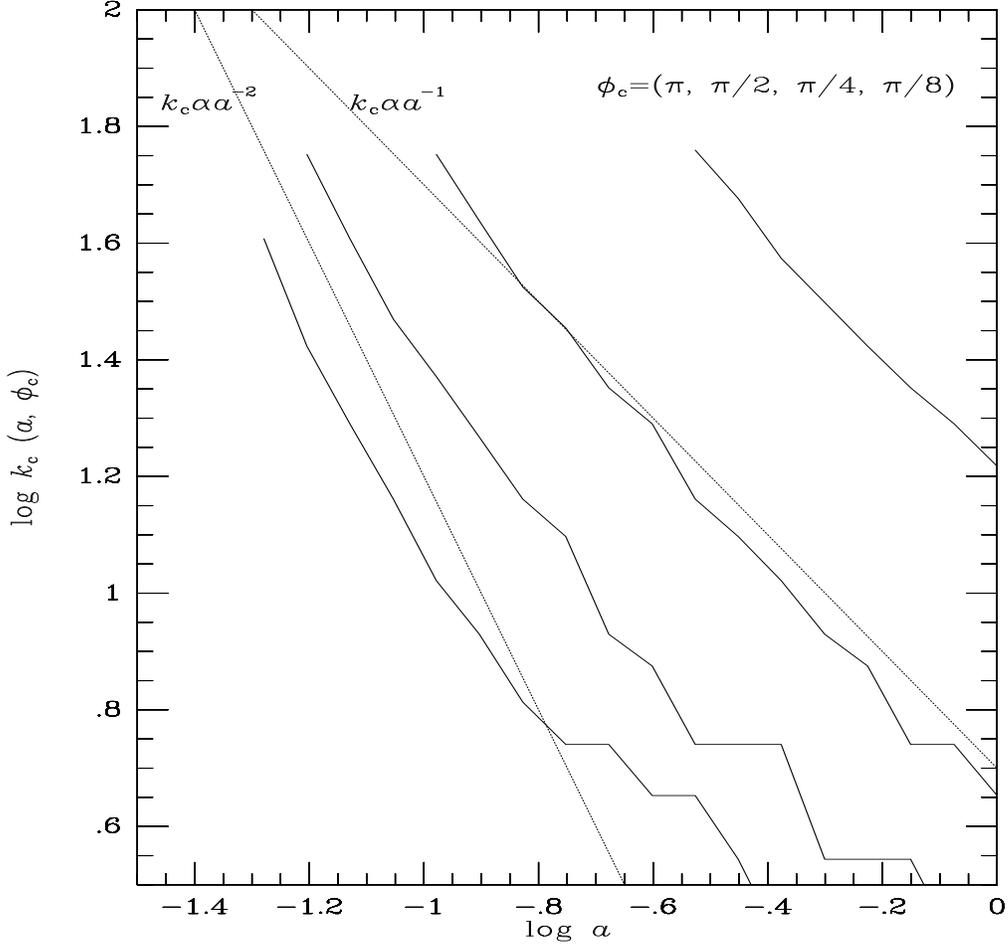}
\label{fig:11}
\end{figure}

\begin{figure}[p]
\vspace*{11.3 cm}
\caption{
%\noindent Fig. 12:
Characteristic scales from the amplitude for $n=-2$.
The scaling of characteristic wavenumber scales
derived from the departure of the amplitude from linear growth, 
$k_c(a,\Delta_c)$ vs. $a$ is shown for 4 different values of $\Delta_c$ (see
equation (\ref{sim2}) for the definition of $k_c(a,\Delta_c)$). 
All the curves have a slope close to the
standard self-similar scaling, $k_c\propto a^{-2}$ at high $k$. For $k$
below about $10$ the slope becomes shallower, probably due to the 
limitation of a finite box. 
}
\includegraphics{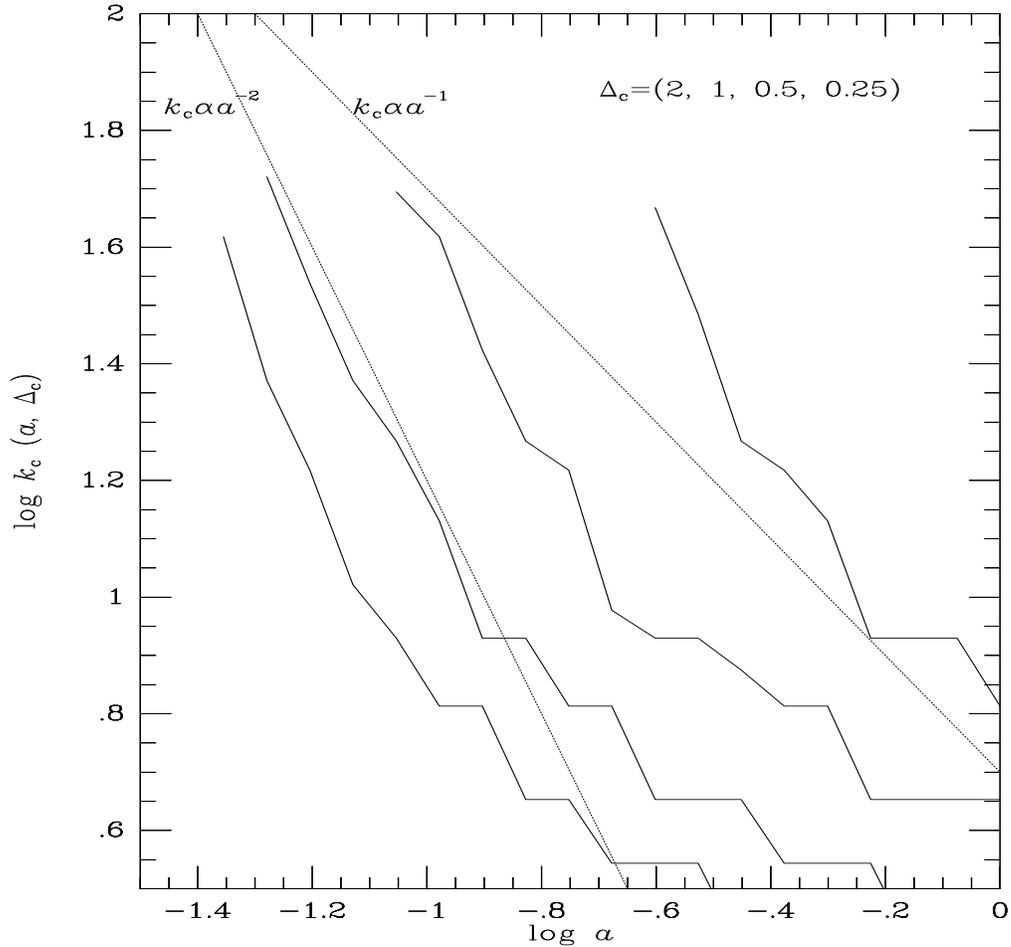}
\label{fig:12}
\end{figure}

\begin{figure}[p]
\vspace*{11.3 cm}
\caption{
%\noindent Fig. 13:
Characteristic scales from the phase for $n=0$.
For $n=0$ the scaling $k_c$ vs. $a$ is shown using the mean phase 
deviation. Notice that in contrast to the $n=-2$ case, here only one 
behavior for all values of $\phi_c$ is evident: the self-similar 
scaling $k_c\propto a^{-2/3}$. 
}
\includegraphics{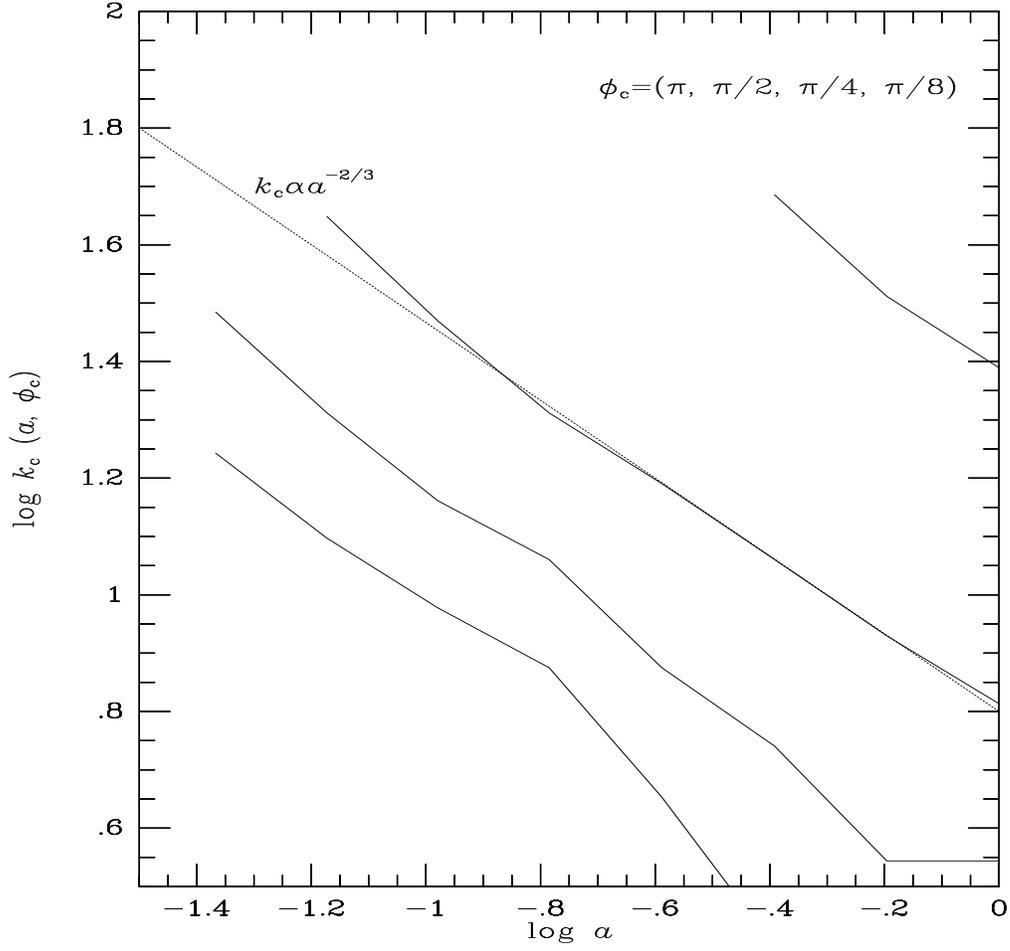}
\label{fig:13}
\end{figure}

\begin{figure}[p]
\vspace*{11.3 cm}
\caption{
%\noindent Fig. 14:
Characteristic scales from the amplitude for $n=0$.
For $n=0$ the scaling $k_c$ vs. $a$ is shown using the mean amplitude. 
Again, as in Figure 13,
%\ref{fig:13}
the standard self-similar scaling is recovered.
}
\includegraphics{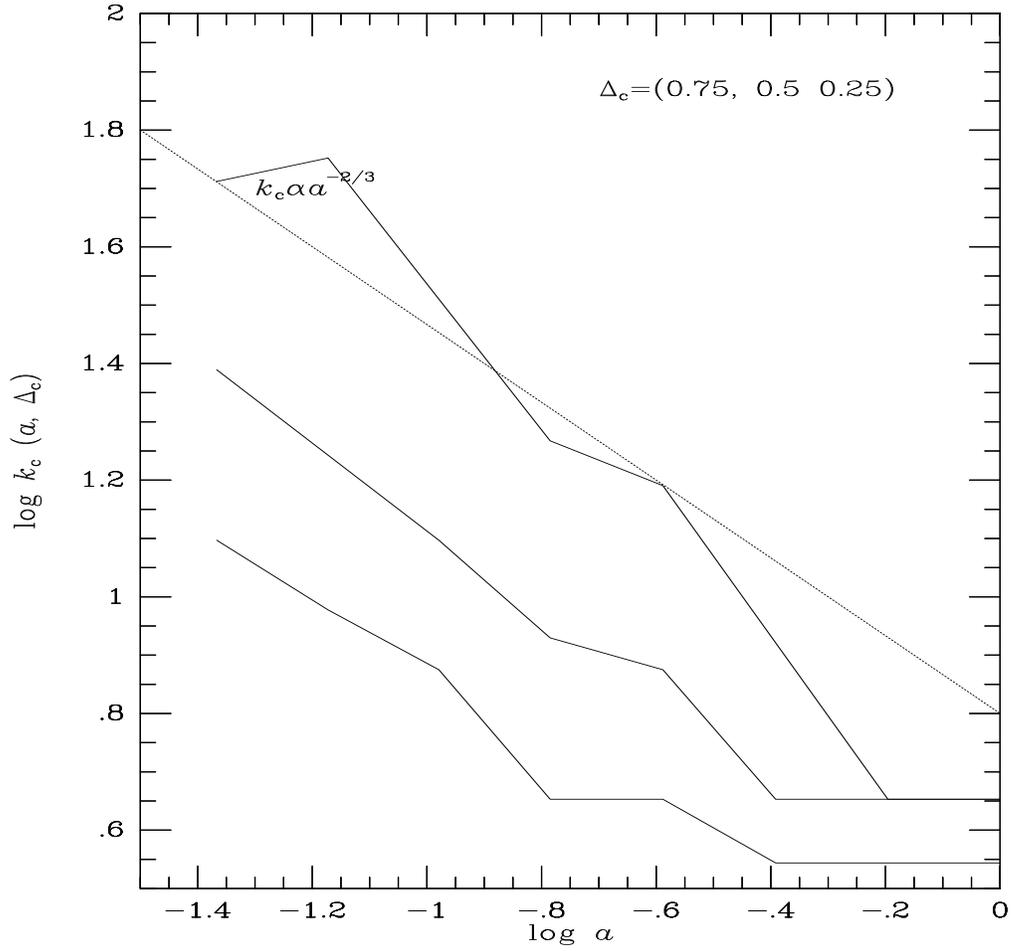}
\label{fig:14}
\end{figure}

\end{document}